\documentstyle[11pt,twoside,jltp,epsbox]{article}
\title{Vortex Nucleation and Array Formation in a Rotating
Bose-Einstein Condensate}

\author{Makoto Tsubota, Kenichi Kasamatsu and Masahito Ueda$^*$}
\address{Department of Physics,Osaka City University,Osaka 558-8585, Japan\\
$^*$Department of Physics, Tokyo Institute of Technology,
Tokyo 152-8551, Japan}

\runninghead{M.Tsubota, K.Kasamatsu and M.Ueda}{Vortex Nucleation
and Array Formation in a Rotating BEC}

\begin{document}

\begin{abstract}
We study the dynamics of vortex lattice formation of a rotating
trapped Bose-Einstein condensate by
numerically solving the two-dimensional Gross-Pitaevskii
equation, and find that the condensate undergoes elliptic deformation,
followed by unstable surface-mode excitations before forming
a quantized vortex lattice.
The dependence of the number of vortices on the rotation
frequency is obtained.

\end{abstract}

\maketitle

\section{INTRODUCTION}
Quantized vortex lattices in the atomic Bose-Einstein Condensates (BECs) have recently been created
experimentally by the groups at ENS\cite{ENS1} and MIT\cite{MIT}.
While several papers \cite{equi} have discussed equiliblium configurations of vortex lattice,
this paper describes the nonequiliblium dynamics towards them by numerically solving
the Gross-Pitaevskii equation(GPE) that governs the time evolution of
the condensate order parameter $\psi({\bf r},t)$:
\begin{equation}
(i - \gamma)\hbar \frac{\partial \psi({\bf r},t)}{\partial t}
= \Bigl[ -\frac{\hbar^2}{2m}\nabla^2+V_{\rm trap}
({\bf r})+ g|\psi({\bf r},t)|^2
-\mu-\Omega L_z \Bigr] \psi({\bf r},t).
\label{GPE}
\end{equation}
Here $g=4\pi \hbar^2 a/m$ is the coupling constant, proportional to the
$^{87}$Rb scattering length $a\approx$5.77nm.
The high anisotropy of the cigar-shaped potential used in the ENS
experiments may permit the two-dimensional analysis.
We thus focus on the two-dimensional dynamics of Eq.(\ref{GPE})
by assuming the trapping potential
\begin{equation}
V_{\rm trap} ({\bf r})= \frac12 m \omega_{\perp} ^2
\{ (1+\epsilon_x)x^2+(1+\epsilon_y)y^2 \},
\label{potential}
\end{equation}
where $\omega_{\perp}= 2\pi \times 219$Hz,
and the parameters $\epsilon_x=0.03$ and
$\epsilon_y=0.09$ describe small deviations of the trap from
the axisymmetry, corresponding to the ENS experiments\cite{ENS1}.
The centrifugal term $-\Omega L_z=i\hbar \Omega (x\partial_y -
y\partial_x)$
appears in a system rotating about the $z$ axis at a frequency $\Omega$.
An important characteristic parameter of the two-dimensional system is
$C=8\pi Na/L$, with the total number $N$ of the condensate
atoms and $L$ the size of the system along the $z$ axis.
The term with $\gamma$ in Eq.\ (\ref{GPE}) introduces the dissipation.

In this paper we investigate the response of the condensate to the sudden
turn-on of the rotation of the potential.
Section 2 describes the scenario of the vortex lattice formation,
following the elliptic deformation of the condensate and the surface-mode
excitations\cite{Tsubota}.
In Sec. 3 we study the dependence of the number of vortices on $\Omega$,
and the critical frequency as a function of $C$.

\section{FORMATION OF VORTEX LATTICE}

We first prepare an equilibrium condensate in a stationary potential.
Figure 1 shows the typical dynamics of the
condensate density $|\psi({\bf r},t)|^2$ after the
potential begins to rotate suddenly with $\Omega = 0.7\omega_{\bot}$\cite{HP}.
The condensate is elongated along the $x$ axis because of the small anisotropy
of $V_{\rm trap}$, and the elliptic cloud oscillates.
Then, the boundary surface of the condensate becomes unstable,
exciting the surface waves which propagate along the surface.
The excitations are likely to occur on the surface
whose curvature is low, i.e.,
parallel to the longer axis of the ellipse.
The ripples on the surface develop into the vortex cores around
which superflow circulates.
Subject to the dissipative vortex dynamics,
some vortices enter the condensate,
eventually forming a vortex lattice. As the vortex lattice is being formed,
the axial symmetry of the condensate is recovered by transferring
angular momentum into quantized vortices.

\begin{figure}
\begin{center}
\epsfile{file=fig1.eps,height=7cm}
\end{center}
\caption{Development of the condensate density $|\psi|^2$ after the trapping potential begins to rotate suddenly with $\Omega=0.7\omega_{\perp}$ for $C=1500$. The time is $t=$0 msec(a), 21 msec(b), 107 msec(c), 114 msec(d), 123 msec(e), and 262 msec(f). The unit for length is $a_{\mathrm HO}=\sqrt{\hbar/2 m \omega_{\perp}}=0.512 \mu$m.}  
\label{fig1} 
\end{figure}

\begin{figure}
\begin{center}
\epsfile{file=fig2.eps,height=7cm}
\end{center}
\caption{Phase profile of $\psi$. The value of the phase varies from $0$ (black) to $2 \pi$ (white). The time of each figure is the same as that in Fig.\ref{fig1} The unit for length is the same as that of Fig.\ref{fig1}.}  
\label{fig2} 
\end{figure}

This peculiar dynamics is understood by investigating
the phase of $\psi({\bf r},t)$ as shown in Fig.\ref{fig2}.
As soon as the rotation of the trapping potential starts,
there come in some branch cuts between the phases 0 and 2$\pi$, whose
ends represent phase defects, i.e. vortices.
When the defects are on the outskirts of the condensate
where the amplitude $|\psi({\bf r},t)|$ is
almost negligible, they neither contribute to the energy nor the angular
momentum of the system.
These defects come into the boundary surface of the condensate within which
the density grows up and the phase defects compete with each other and 
induce the above surface waves due to interference.
Then the selection of the defects starts, because their further invasion
costs the energy and the angular momentum.
As is well known in the study of rotating superfluid helium \cite{rota},
the rotating drive pulls vortices
into the rotation axis, while the intervortex repulsive interaction tends to push them
apart; their competition yields a vortex lattice whose vortex density
depends on the rotation frequency.
In our system, some vortices enter the
condensate and form a lattice dependent on
$\Omega$, while excessive vortices are repelled and escape to the outside.
Remarkably, the phase profile of Fig.2(f) reveals that the repelled vortices
also form a lattice on the outskirts of the condensate. Since they cannot
be seen in the corresponding density profile of Fig.1(f), they may be called
``ghost" vortices.

The distortion of the condensate to an elliptic
cloud was theoretically studied
by Recati et al.\cite{Recati}.
Assuming the quadrupolar velocity field
${\bf v}({\bf r}) = \alpha \nabla (xy)$,
they obtained the distortion parameter
$\alpha=\Omega (R_x^2 - R_y^2)/(R_x^2 + R_y^2)$ in the steady states as a
function of $\Omega$, where $R_x$ and $R_y$ are the sizes of the
condensate along each direction.
Madison et al. observed that, after the rotation
of $\Omega = 0.7\omega_{\bot}$
starts suddenly, $\alpha/\omega_{\perp}$ oscillates during a few
hundred milliseconds and then falls abruptly to a value below 0.1 when vortices
enter the condensate from its boundary surface\cite{ENS2}.
Figure 3 shows the oscillation of $\alpha$ and the increase of the angular
momentum $\ell_z/\hbar$ per atom in our dynamics of Fig.1.
This figure closely resembles Fig.3 of Ref. 7,
and our scenario is consistent with the experimental results.

\begin{figure}
\begin{center}
\epsfile{file=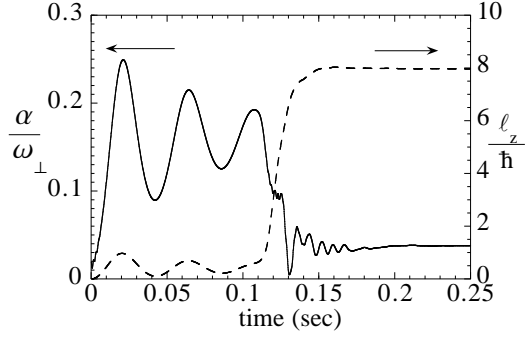,height=5cm}
\end{center}
\caption{Time evolution of the distortion parameter $\alpha$ (solid line) and the angular momentum $\ell_{z}/\hbar$ per atom (dashed line) in the dynamics of Fig.1.}  
\label{fig3} 
\end{figure}

The dispersion relation of the surface waves\cite{Khawaja} of a rotating
condensate is given by $\omega^2=R(\omega_{\bot}^2-\Omega^2)k$ with the
condensate size $R$.
The dependence of $\omega$ on $R$ shows that the surface with lower curvature
is excited more easily, which is clearly seen in Fig.1.
The group velocity $d\omega/dk= \sqrt{R(\omega_{\bot}^2-\Omega^2)}/(2\sqrt{k})$
agrees with the propagation velocity of the surface wave in our simulation.

\section{CRITICAL FREQUENCIES AND MULTIPLE-VORTICES LATTICES}
We have examined the critical frequency at which vortices can enter. Figure 4(a) shows the dependence of the number of vortices on the frequency $\Omega/\omega_{\perp}$, for $C=$250, 450 and 1500. For $C=450$ corresponding to the experimental condition \cite{ENS1}, we found that only when $\Omega$ exceeds $\Omega_{c1} \simeq 0.56\omega_{\perp}$ vortices can enter the condensate and form a lattice. This critical frequency is closer to the observed value $\Omega_{\rm nuc} = 0.65\omega_{\perp}$, than the values obtained in previous literature \cite{cri}. The increase in $C$ reduces the critical frequency, and stabilizes the lattices of more vortices for the same frequency, The increase in $\Omega$ causes lattices of multiple vortices; figure 4(b) shows a lattice of 44 vortices for $C=$1500.

The critical frequencies have been studied from thermodynamic or stability arguments\cite{cri,Isoshima}, but they are generally much smaller than the observed value. It should be noted that these critical frequencies only give the necessary condition which enables a vortex to exist stably at the center of the trap. Actually, vortices should be nucleated at the boundaries and come into the condensate; the condition of $\Omega$ which realizes such nonlinear dynamics may be generally different from that obtained from the stability arguments. Isoshima and Machida examined the local instability of nonvortex states toward vortex states within the Bogoliubov theory\cite{Isoshima}. They note that this local instability could give higher critical frequencies than the global instability which comes from comparing their total energy. From Fig.4(a) the critical frequencies are 0.58, 0.56 and 0.45, respectively for $C=$250, 450, and 1500. They agree with the critical frequencies 0.57, 0.52 and 0.41 for the same values of $C$ obtained from the local stability analysis \cite{Isoshima}.

\begin{figure}
\begin{center}
\epsfile{file=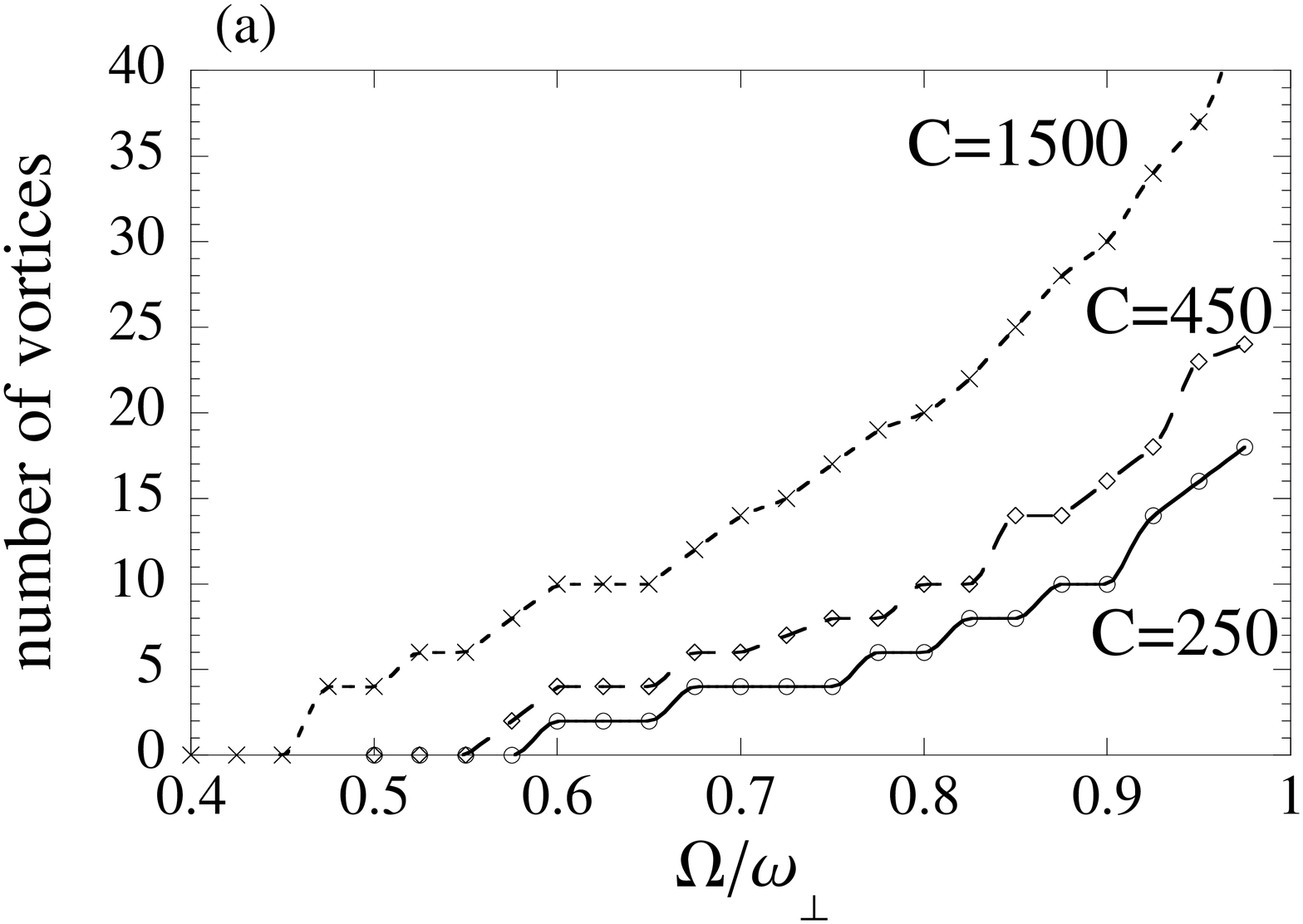,height=4.3cm}
\epsfile{file=fig4b.eps,height=4.3cm}
\end{center}
\caption{(a) Number of vortices versus $\Omega/\omega_{\perp}$, for $C=$250, 450 and 1500. (b) Lattice of 44 vortices for $C=$1500 and $\Omega/\omega_{\perp}=$ 0.975.}  
\label{N-Omega} 
\end{figure}

\section{CONCLUSIONS}
We have studied the dynamics of vortex lattice formation of
a rotating trapped BEC by numerically solving the two-dimensional GPE, and
obtained the following physical picture.
When the trapping potential begins to rotate
at a sufficiently fast frequency,
the condensate is distorted to an elliptic shape and oscillates.
Then its boundary surface becomes unstable, exciting surface waves.
The origin of these ripples
is identified to be the interference of violent phase fluctuations
that occur on the outskirts of the condensate;
and some of the surface ripples  develop
into the vortex cores which then form a vortex lattice.
The dynamics is concerned with the ``ghost" vortices which are the
phase defects outside the condensate.

\section*{ACKNOWLEDGMENTS}
The authors thank K.Machida for useful discussions.
MT acknowledges support by a Grant-in-Aid for Scientific Research
(Grant No.12640357) by Japan Society for the Promotion of Science.
MU acknowledges support
by a Grant-in-Aid for Scientific Research
(Grant No.11216204) by the Ministry of Education, Culture, Sports,
Science and Technology of Japan, and by the Toray Science Foundation.

\end{document}